\documentclass[preprint,12pt]{elsarticle}

\usepackage[T1]{fontenc}
\usepackage{graphicx}       
\usepackage{rotating}
\usepackage{booktabs}
\usepackage{tabularx}
\usepackage{adjustbox}
\usepackage{subcaption}     
\usepackage{makecell}
\usepackage{multirow}
\usepackage{array}
\usepackage{enumitem}
\usepackage{ragged2e}        
\usepackage{threeparttable}  
\usepackage{microtype}   
\usepackage[colorlinks=true, linkcolor=blue, citecolor=blue, urlcolor=blue]{hyperref}

\newcolumntype{P}[1]{>{\raggedright\arraybackslash}p{#1}}
\newcolumntype{C}[1]{>{\centering\arraybackslash}p{#1}}
\setlist[itemize]{nosep,left=0pt,topsep=0pt,partopsep=0pt}
\biboptions{sort&compress}

\begin{document}

\title{Everyone Needs AIR: An Agnostic Incident Reporting Framework for Cybersecurity in Operational Technology}

\author[napier]{Nubio Vidal\corref{cor1}}
\ead{n.vidal@napier.ac.uk}

\author[napier]{Naghmeh Moradpoor}
\ead{n.moradpoor@napier.ac.uk}

\author[napier]{Leandros Maglaras}
\ead{l.maglaras2@napier.ac.uk}

\cortext[cor1]{Corresponding author}

\address[napier]{School of Computing, Engineering \& the Built Environment, Edinburgh Napier University, Edinburgh EH10 5DT, UK}

\begin{abstract}
Operational technology (OT) networks are increasingly coupled with information technology (IT), expanding the attack surface and complicating incident response. Although OT standards emphasise incident reporting and evidence preservation, they do not specify what data to capture during an incident, which hinders coordination across stakeholders. In contrast, IT guidance defines reporting content but does not address OT constraints. This paper presents the Agnostic Incident Reporting (AIR) framework for live OT incident reporting. AIR comprises 25 elements organised into seven groups to capture incident context, chronology, impacts, and actions, tailored to technical, managerial, and regulatory needs. We evaluate AIR by mapping it to major OT standards, defining activation points for integration and triggering established OT frameworks, and then retrospectively applying it to the 2015 Ukrainian distribution grid incident. The evaluation indicates that AIR translates high-level requirements into concrete fields, overlays existing frameworks without vendor dependence, and can support situational awareness and communication during response. AIR offers a basis for standardising live OT incident reporting while supporting technical coordination and regulatory alignment.
\end{abstract}


\maketitle

\section{Introduction} \label{intro}

Operational technology (OT) is the foundation for modern industry and critical infrastructure, but its expanding scope and increasing interconnection with information technology (IT) broaden the attack surface. On 23 December 2015, attackers penetrated three oblenergos, Ukrainian regional power distribution entities. Using the BlackEnergy3 malware, they stole credentials from the Supervisory Control and Data Acquisition (SCADA) network, opened 30 breakers, and cut power to approximately 225,000 customers, triggering a manual restoration effort that lasted up to six hours \cite{NCCIC-ICS-article,Booz2019}. In addition, the attackers used tactics to disrupt collaboration and impede mitigation efforts: they corrupted the firmware of the remote terminal unit, flooded the call centres with telephone denial of service attacks, disabled uninterruptible power supply units (UPS units) and deployed a secondary malware (KillDisk) targeting corporate and industrial control system networks \cite{NCCIC-ICS-article,Whitehead2017}. These actions limited response teams and real-time situational awareness.

Whitehead et al. \cite{Whitehead2017} reported the absence of correlated syslogs, packet captures, and intrusion detection alerts as a factor delaying incident diagnosis. This problem reflects a broader challenge in OT environments: the lack of structured and standardised approach to incident disclosure. Reports about the 2015 Ukrainian incident illustrate how improved information exchange could have enhanced resilience and more effective real-time response \cite{SANS2016}.

OT-specific standards, such as the ISA/IEC 62443 and the NERC-CIP series, suggest reporting cybersecurity incidents and forensic evidence preservation \cite{IEC62443,NIST80082r3,NERCCIP}. Yet, these standards do not explicitly propose the elements to be disclosed. This vague approach to reports impedes collaboration during live incidents and impacting operational continuity. Organisations customise reporting elements through service-level agreements (SLAs) to address this absence, resulting in fragmented, inconsistent, and incomplete exchanges during incidents \cite{HomelandDefense}. The lack of a typical structure adds uncertainty and increases the cost of post-incident analysis because evidence can be overlooked.

In contrast, IT-focused standards, like the ISO/IEC 27035 series and NIST SP 800‑61 (rev. 3), provide detailed guidance on incident reporting content, including indicators of compromise (IOCs), affected assets, severity levels, timestamps, and mitigation actions \cite{ISO27035,NIST80061r3}. These standards encourage continuous reporting throughout the detection, containment, and recovery phases, rather than limiting it to post-incident summaries. Still, they overlook the aspects of OT environments, for instance, documenting deviations in process safety parameters (e.g., pressure or voltage anomalies) and capturing the operational behaviour of industrial control systems (ICS).

The OT cybersecurity frameworks proposed in the literature prioritise incident identification and response, also failing to tackle the reporting problem, as detailed in subsection \ref{lit_fram}. A recent stream of thought suggests automated playbooks applying predefined vendor remediation actions, as discussed in \cite{empl2024ics}. Nevertheless, they can be ineffective against unanticipated threats \cite{Khadidos2025CyberSentry}.

This study proposes AIR, an Agnostic Incident Reporting framework tailored to ongoing OT cybersecurity incidents, to address this identified critical gap in the disclosure process. AIR is a modular framework that draws on the structure and best practices of IT standards \cite{ISO27035,NIST80061r3}, while adapting them to the operational constraints of industrial systems. AIR is agnostic because it does not assume alignment with any specific OT security architecture. AIR complements the existing OT standards \cite{IEC62443,NIST80082r3,NERCCIP} by providing a baseline reporting format that these standards currently lack. It is designed to harmonise reporting fields, terminology, and timing across multi-stakeholder OT environments. The adaptation of IT standards for OT use is a recognised practice (see \cite{NIST80082r3}), and AIR advances this approach by offering an adaptable foundation for organisations to extend their reports based on their operational needs and comply with regulatory obligations.

The core contribution of AIR lies in its operationalisation of structured reporting as a central element of OT incident response, an underdeveloped dimension in the OT cybersecurity literature. The implementation of a structured report also incentivises collaboration and operational continuity, which are vital for critical infrastructure. AIR enables organisations to align incident reporting with SLAs and regulatory expectations, regardless of their current cybersecurity maturity, by providing an extensible, technology-agnostic template. It offers regulators and standards bodies a blueprint for future enhancements to OT-specific reporting guidance.

The rest of this paper is structured as follows. Section \ref{rev} reviews the literature and OT/IT standards. Section \ref{AIR} details our proposition: the AIR framework. In Section \ref{AIR_practice}, we design exercises to assess AIR's compliance with OT standards, scalability, and operational applicability. Section \ref{final} offers concluding remarks, outlines limitations, and suggests future work.

\section{Background Review} \label{rev}

OT environments face cyber risk as they interconnect with IT networks and partners. Frameworks and standards seek to secure OT assets, coordinate incident response and assign organisational roles. Yet, they differ in coverage, technical depth and fitness for real-time, multi-stakeholder incidents. This section (i) reviews OT-specific methodological frameworks (sec. \ref{lit_fram}); (ii) analyses how key OT standards treat incident reporting and coordination (sec. \ref{ot_stand}); and (iii) evaluates mature IT-security standards that could underpin a neutral incident-reporting framework (sec. \ref{it_stand}).

\subsection{OT Frameworks in Literature} \label{lit_fram}

Past frameworks address the detection, analysis, and mitigation of cybersecurity incidents in OT, differing in technical orientation, purpose, and operational depth, and varying from adaptive response systems (e.g., \cite{Khadidos2025CyberSentry}) to recovery planning (e.g., \cite{onwubiko2024recovery}). In Table \ref{tab:lit_fram}, we categorise those frameworks into groups, further described in the following subsections, to summarise their focus and limitations. We also added an explanation about the benefits of adding the ``AIR layer'' in those groups (refer to section \ref{ex_ot_adapt} for individual examinations), given that they overlooked the incident report. 

\begin{table}[ht]
\centering
\caption{Groups of OT literature frameworks, limitations, and AIR benefits}\label{tab:lit_fram}
\begingroup
\setlength{\tabcolsep}{4pt}
\renewcommand{\arraystretch}{1.1}
\resizebox{\textwidth}{!}{
\begin{tabular}{p{2.2cm} p{3.4cm} p{3.0cm} p{4.2cm} p{4.0cm}}
\toprule
\textbf{Group} & \textbf{Examples} & \textbf{Focus} & \textbf{Limitation} & \textbf{AIR benefit} \\
\midrule
Detection and adaptive response &
CyberSentry \cite{Khadidos2025CyberSentry}; DSACR \cite{ma2024dsacr}; POROS \cite{iaiani2021poros}; PRAETORIAN \cite{Papadopulos2024PRAETORIAN}; XA4AS \cite{seid2024xa4as} &
Real-time or adaptive threat detection and mitigation &
Engines remain site-centred; limited ability to escalate or share data across organisational boundaries &
Compiles pre-processed incident fields into a concise report with no extra load on real-time loops \\
\addlinespace
Testing and resilience architectures &
CROF \cite{onwubiko2024recovery}; PRESeT \cite{schaeffer2014resilience}; TiDICS \cite{staves2024framework} &
Testing, recovery planning, and resilience integration &
No regulator-ready log format; short maintenance windows do not support cross-vendor evidence exchange; weak multi-party handling &
Produces regulator-ready records and supports coordinated, multi-party incident handling \\
\bottomrule
\end{tabular}
}
\endgroup
\end{table}

\subsubsection{Detection and adaptive response frameworks}

CyberSentry \cite{Khadidos2025CyberSentry} utilises deep learning techniques, through Tri-Fusion Net, and optimisation methods (from Parrot-Levy Blend Optimisation) to improve incident detection and response. The model relies on SCADA capabilities to share information. Ma et al. \cite{ma2024dsacr} present DSACR, which is an adaptive cyber resilience framework. DSACR defensive strategies considering the severity and operational impact of the detected incidents. Although balancing security, performance, and efficiency, DSACR relies on real-time OT/IT telemetry, reducing its cross-site transferability. 

POROS \cite{iaiani2021poros} is a nine-step framework to identify security events resulting from the malicious manipulation of OT systems. It allows third-party participation but lacks an inter-organisational escalation process, thereby limiting its applicability. Papadopoulos et al. \cite{Papadopulos2024PRAETORIAN} design PRAETORIAN that covers the cyber and physical aspects of critical infrastructure. PRAETORIAN primarily relies on sensors (physical domain) and information gathered from security information and event management tools (cyber domain). Although it has a tool enabling information sharing across first responders, the information exemplified there is not an actual list, nor the specific set if it were an OT cybersecurity incident, making its application reliant on ad hoc agreements. XA4AS \cite{seid2024xa4as} is a predictive-control framework that dynamically alters defence policies through behavioural monitoring. Although control actions are distributed to edge nodes, XA4AS relies on a single, globally agreed-upon optimisation goal, considering a set of predetermined goals.

\subsubsection{Testing and resilience architecture frameworks}

The Cyber Recovery Operational Framework (CROF) \cite{onwubiko2024recovery} adapts NIST guidance for cyber recovery, prompting internal coordination but not multi-party recovery strategies in incident management. PRESeT \cite{schaeffer2014resilience} sets out a resilience architecture using reusable patterns to align detection and mitigation strategies. It supports distributed anomaly detection and policy-based adaptation. However, it omits a regulator-direct reporting hook, diverging from the reality of industrial settings. Following the same pattern is TiDICS \cite{staves2024framework}, a testing framework that integrates safety and operational risk into scoping adversary-centric security assessments. Although methodologically rigorous, it has a scalability issue related to test-window length. 

Whereas each example in the preceding groups enhance cyber-resilience for users, they fail to account for the complexity of federated OT environments, the organisational standard, where multiple parties share control and visibility. As a result, frameworks lack mechanisms for structured incident reporting. Addressing this gap surpasses ad hoc agreements, grounding them in standards that define shared information settings. The proposed AIR framework offers organisations a shareable baseline of pre-processed data concerning OT cybersecurity incident information, rooting SLAs regarding the input needed to coordinate incident management while remaining compliant with OT standards. This relation is further discussed in Section \ref{ex_ot_adapt}.

\subsection{OT standards} \label{ot_stand}

\begin{figure}[ht]
 \centering
 \includegraphics[width=0.7\textwidth]{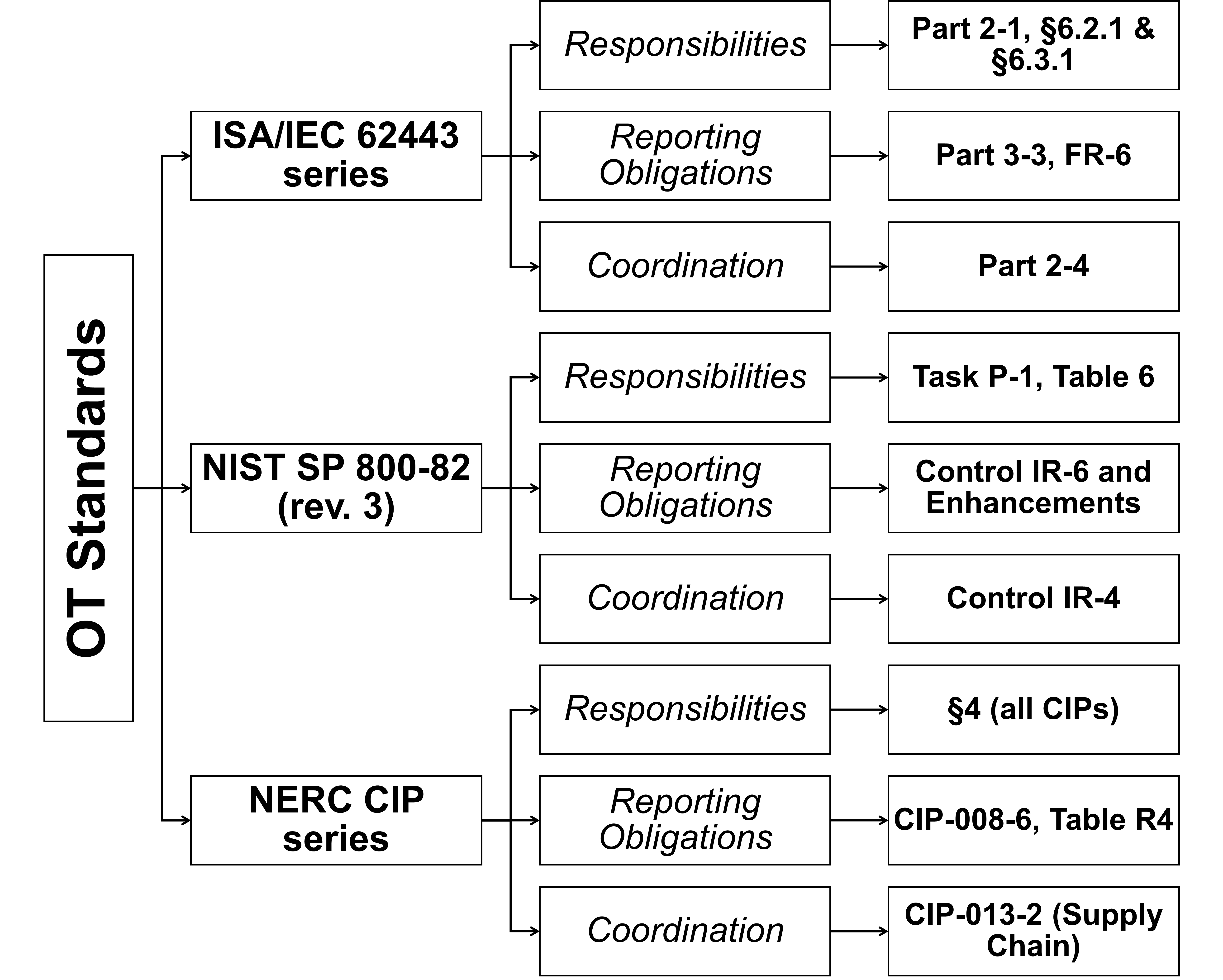}
 \caption{OT standards: responsibilities, reporting and coordination}
 \label{fig:ot_stand}
\end{figure}

Surveyed frameworks did not dictate the exchange of field-level data during an ongoing incident, while OT standards indirectly address the topic through incident-response controls. Because regulators, contracts and audits reference these texts, organisations align processes with external expectations. This alignment can duplicate reports and create incompatible data fields or audit gaps in multi-stakeholder settings. Fig. \ref{fig:ot_stand} pinpoints where OT standards refer to responsibilities, define the reporting process and frame coordination.

The ISA/IEC 62443 series \cite{IEC62443} consists of several parts. Part 2-1 emphasises risk-based controls and formal role assignment across the asset-owner ecosystem. Part 3-3 (Foundational requirement 6, FR-6) scales reporting from passive log retention to near-real-time push notifications, but omits a canonical report template. Cross-organisation exchange remains an ad hoc matter governed by non-disclosure agreements (NDAs) or SLAs (Part 2-4).

NIST SP 800-82 (rev. 3) \cite{NIST80082r3} is built on the NIST Risk Management Framework. The standard lists OT-specific roles and then directs users to prompt incident reporting under control IR-6. While its enhancements encourage automated feeds and supply-chain visibility, the standard leaves field definitions to local policy. Coordination (in IR-4) is expected from external vendors, integrators, and suppliers, despite no workflow prescription.

The North American Electric Reliability Corporation (NERC) released the Critical Infrastructure Protection (CIP) series \cite{NERCCIP}. NERC CIP was designed for Bulk Electric Systems but has been extended to other critical infrastructure. Responsibilities depend on the scope and are outlined in Section 4 of each CIP. CIP-008-6 requires entities to notify the Electricity Information Sharing and Analysis Center (E-ISAC) and the United States National Cybersecurity and Communications Integration Center (NCCIC), if falls under U.S. jurisdiction, within one hour, providing details on the functional impact, attack vector, and intrusion depth (Table R4). CIP-013-2 pushes coordination clauses into vendor contracts, but routine joint handling procedures are limited.

National guidances, like from Singapore (CSA, CII Code of Practice\footnote{Refer to: \url{https://isomer-user-content.by.gov.sg/36/2df750a7-a3bc-4d77-a492-d64f0ff4db5a/CCoP---Second-Edition_Revision-One.pdf}}), Saudi Arabia (NCA, OT Cybersecurity Controls\footnote{Refer to: \url{https://nca.gov.sa/ar/otcc_en.pdf}}), France (ANSSI, Managing Cybersecurity for ICS\footnote{Refer to: \url{https://cyber.gouv.fr/sites/default/files/2014/01/Managing_Cybe_for_ICS_EN.pdf}}), Japan (METI, Cyber/Physical Security Framework\footnote{Refer to: \url{https://www.meti.go.jp/policy/netsecurity/wg1/CPSF_ver1.0_eng.pdf}}), or the United Kingdom (Product Security and Telecommunications Infrastructure Act\footnote{Refer to: \url{https://www.legislation.gov.uk/ukpga/2022/46/contents}}), provide governance guidance but do not specify structured reporting frameworks for ongoing cybersecurity incident in OT contexts.

The OT standards provide several guidelines for cybersecurity incident management, but they fall short in reporting cybersecurity incidents, particularly while they are ongoing, when multidimensional cooperation is required. To handle this, we proposed the AIR framework, which helps organisations at various cybersecurity maturity levels overcome customised SLAs. The perceived persistent gap in OT‐specific standards motivates the turn to IT alternatives in the search for more granular models of incident reporting.

\subsection{IT standards}\label{it_stand}

\begin{figure}[ht]
 \centering
 \includegraphics[width=0.7\textwidth]{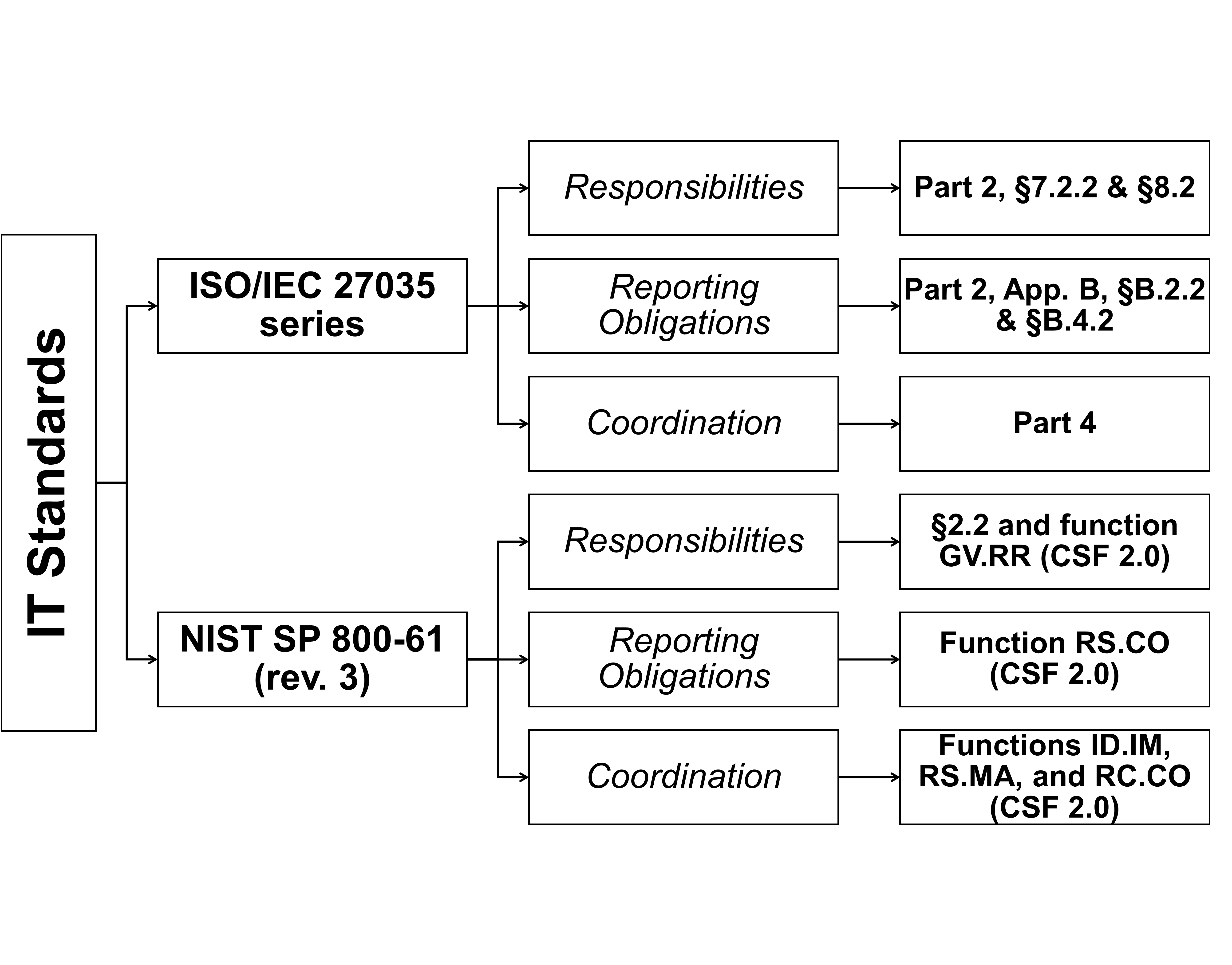}
 \caption{IT standards: responsibilities, reporting and coordination}
 \label{fig:it_stand}
\end{figure}

OT frameworks give scant advice on what to exchange during an ongoing incident. Mature IT standards may serve as a reference, primarily to outline the data that should be shared in the event of a cybersecurity incident. Fig. \ref{fig:it_stand} identifies the location of information regarding responsibilities, incident reporting, and coordination within two widely adopted IT standards.

The ISO/IEC 27035 series \cite{ISO27035} offers some of the most granular guidance among mainstream IT standards. It specifies incident-content elements, provides checklists for consistent documentation, and details how separate organisations can share information and coordinate actions. Nonetheless, the series remains generic and suggests that sector-specific tailoring is necessary before it can function as a real-time and inter-stakeholder reporting scheme.

NIST SP 800-61 (rev. 3) \cite{NIST80061r3} defines a four-phase incident-handling lifecycle and assigns duties to incident handlers, management, and external partners. Its alignment with the NIST Cybersecurity Framework (e.g., RS.CO on communication) encourages timely information flow, but avoids prescribing a field-level template, which leaves content definitions to SLAs and limits interoperability in multi-party incidents.

Compared to OT standards, IT alternatives, especially the ISO/IEC 27035 series, offer more mature approaches to incident communication. Building on this maturity, the following Section introduces AIR, an Agnostic Incident Reporting framework tailored for OT contexts. 

\section{Agnostic Incident Reporting (AIR) framework for cybersecurity in operational technology} \label{AIR}

Our AIR framework adapts selected elements from the ISO/IEC 27035 series and NIST SP 800-61 (rev. 3). It translates them into a format that reflects the technical and organisational realities of OT environments, grounding the disclosure process in SLAs and complying with OT standards.

\subsection{Overview}

\begin{figure}[ht]
 \centering
 \includegraphics[width=\textwidth]{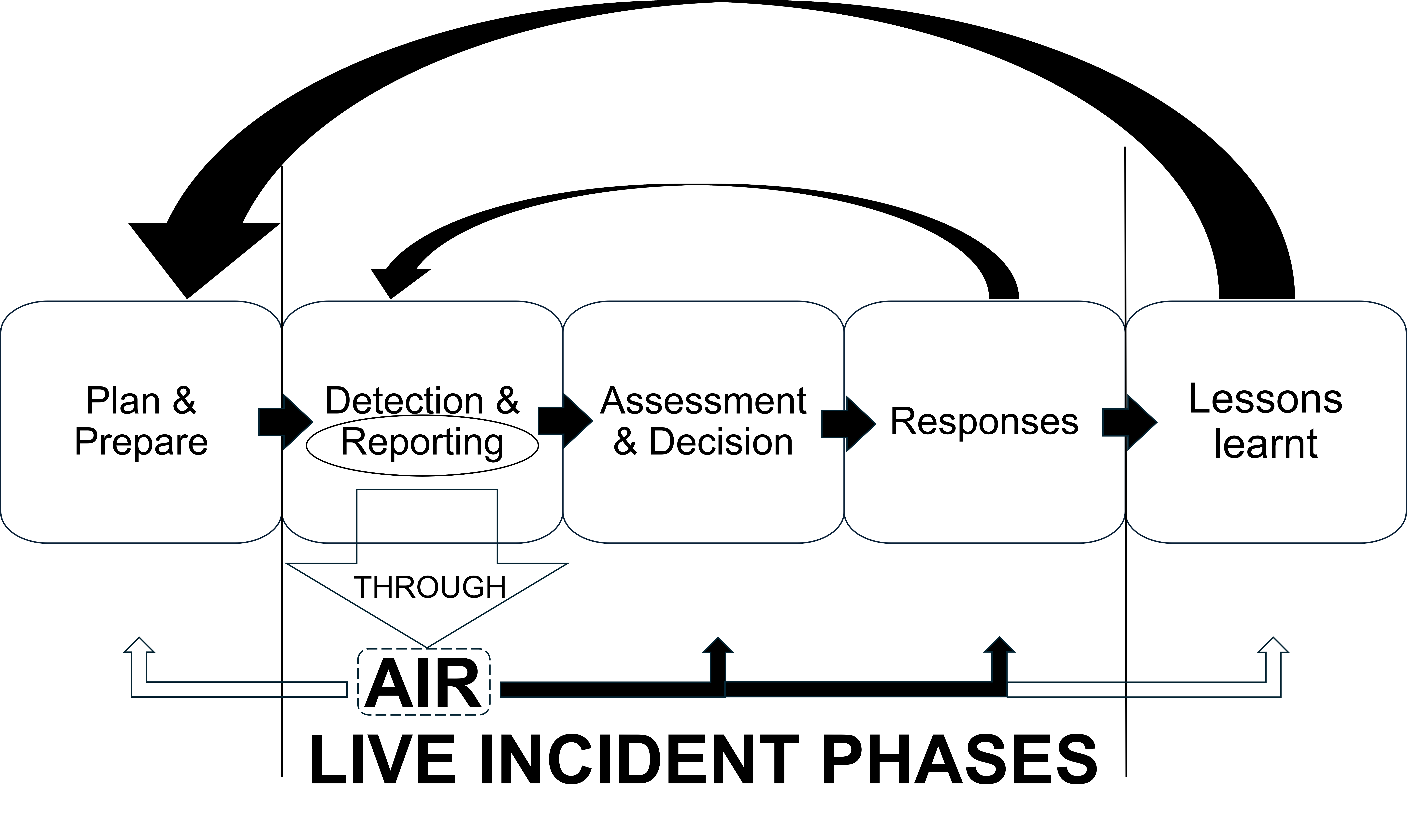}
 \caption{AIR in the ISO 5-phase model incident management lifecycle}
 \label{fig:AIR_lifecycle}
\end{figure}

In line with the ISO/IEC 27035 series, AIR enhances incident response by providing structured guidance for ongoing OT incidents and follows the five-phase model in \cite[Part 1, Section 5.1]{ISO27035}. By contrast, NIST SP~800-61 (rev. 3) \cite{NIST80061r3} organises guidance by the NIST CSF 2.0 Functions and does not prescribe a field-level reporting template. As shown in Fig. \ref{fig:AIR_lifecycle}, AIR integrates with Phase 2 (Detection \& Reporting) and its structured baseline also supports Phase 3 (Assessment \& Decision) and Phase 4 (Response).

While AIR operates during the active phases of an incident, its effects extend beyond them. The resulting detailed documentation reduces friction during the Lessons learned phase. These insights also reinforce the Plan and prepare phase, strengthening organisational cybersecurity maturity. The white angled arrows in Fig. \ref{fig:AIR_lifecycle} indicate these relations.

\subsection{AIR Global Structure and OT Relevance} 

\begin{sidewaysfigure}
  \centering
  \includegraphics[width=\textwidth]{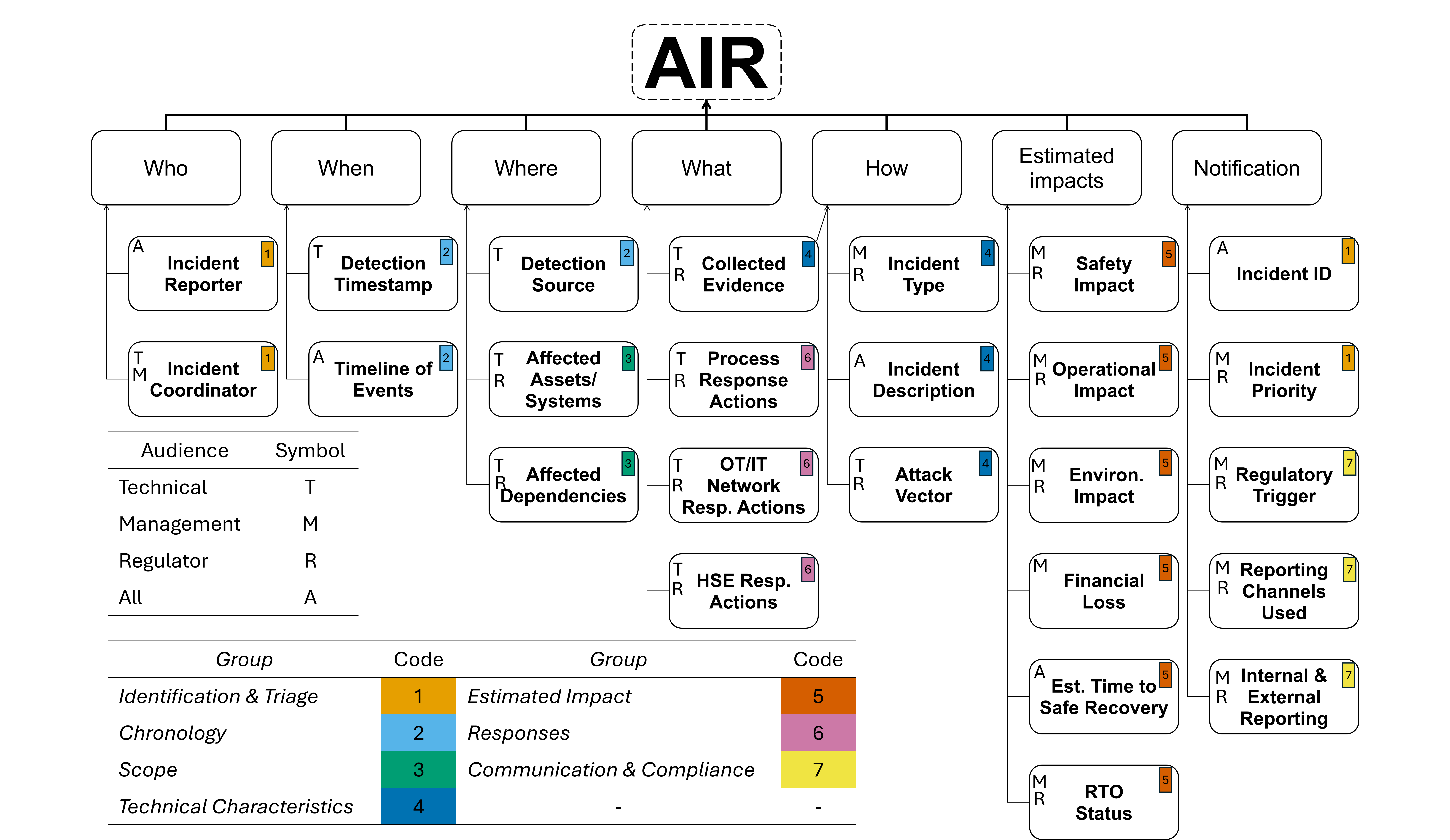}
  \caption{AIR elements and proposed audiences}
  \label{fig:AIR_elem_aud}
\end{sidewaysfigure}

AIR comprises seven groups (italicised) comprising 25 elements (in bold). The framework aligns with the incident-handling loop in Fig. \ref{fig:AIR_lifecycle} and is depicted in Fig. \ref{fig:AIR_elem_aud}. Most frameworks pre-process many of these elements. AIR can compile them with minimal overhead. The grouping is didactic since, during ongoing OT incidents, response teams capture several elements in parallel.

\textit{Identification \& Triage} and \textit{Chronology} anchor the record. Once the incident is triggered, a unique \textbf{Incident ID} follows the case across organisations. \textbf{Incident Priority} collects the result from the plant risk matrix (see \cite[Part 3-2, Annex B]{IEC62443}). Named contacts for the \textbf{Incident Reporter} and the \textbf{Incident Coordinator}, including their role and responsibilities, ensure accountable actions. Time is fixed with a \textbf{Detection Timestamp} (UTC) and a \textbf{Detection Source} such as a SOC console or an operator call. A running \textbf{Timeline of Events} that orders observations and actions. 

The \textit{Scope} group captures what is directly or indirectly affected: \textbf{Affected Assets/Systems} (OT/ICS elements confirmed as compromised) and \textbf{Affected Dependencies} (for example, safety controllers or IT gateways), helping analysts detect lateral movement. Turning to the threat, \textit{Technical Characteristics} record the \textbf{Incident Type} using a recognised taxonomy (in \cite[Section C.1]{NIST80082r3}), \textbf{Attack Vector}, and short \textbf{Incident Description} in plain language. \textbf{Collected Evidence} lists preserved artefacts, such as logs and memory dumps, with chain-of-custody notes.

In the \textit{Estimated Impact} group, we record \textbf{Safety Impact}, \textbf{Operational Impact}, \textbf{Environmental Impact} and \textbf{Financial Loss} for effects that are expected or observed. The \textbf{Estimated Time to Safe Recovery} projects a time window to a safe state, and \textbf{RTO Status} shows progress against a predefined recovery time objective (RTO).

Within \textit{Responses}, actions are: \textbf{Process Response Actions} (plant interventions, e.g. valve isolation), \textbf{OT/IT Network Response Actions} (such as segmentation and patching), and \textbf{HSE Response Actions} (steps covering health, safety and environmental responses). Recording mitigation actions in real time aids coordination and subsequent lessons learned.

\textit{Communication and Compliance} sets out reporting and notification. The \textbf{Regulatory Trigger} identifies the legal basis, \textbf{Reporting Channels Used} lists the sharing mechanisms already employed, and \textbf{Internal and External Reporting} tracks the reporting status by stakeholders.

Sensitive data may be shared with partners under NDAs, while regulators operate under statutory powers that can override such agreements in some jurisdictions. These combined elements provide a standardised template that supports quick sharing and preserves detail, clarity and regulatory compliance.

OT operations must balance personnel safety, service continuity, and cross-party coordination \cite{ma2024dsacr,seid2024xa4as,staves2024framework}. Fig. \ref{fig:AIR_elem_aud} shows that the answers to the elements map to who is involved, when events occurred, where compromise happened, how the intrusion unfolded, what actions were taken, estimated impacts, and notification fields. \textbf{Collected Evidence} is the only element connecting to two questions: what happened and how the incident occurred.

AIR defines information audiences, aligning with the IT ``need-to-know'' principle \cite[Part 2, p. 12]{ISO27035}; among surveyed frameworks, only PRAETORIAN makes this explicit \cite{Papadopulos2024PRAETORIAN}. AIR adopts the principle and defines audience-specific datasets so that parties receive only essential information, while technical teams, management, and regulators can collaborate in parallel.

\subsection{AIR elements: IT standards mapping to OT relevance} 

Previously, we noted that AIR draws on IT standards. Table \ref{tab:air_mapping_it} maps each AIR element to one or both of the mature IT security standards examined here, highlighting their use as a starting point for OT reporting. The ISO/IEC 27035 links are taken from the list of ``recorded items for information security incidents'' \cite[Part 2, Section B.2.2]{ISO27035}. The NIST mappings follow the CSF 2.0 references provided in \cite{NIST80061r3}, as summarised in its mapping tables.

\begin{table}[htbp]
\centering
\caption{Mapping of AIR elements to IT standards}
\label{tab:air_mapping_it}
\begin{threeparttable}
\begingroup
\footnotesize
\setlength{\tabcolsep}{2.5pt}         
\renewcommand{\arraystretch}{1.02}    
\setlength{\aboverulesep}{0pt}        
\setlength{\belowrulesep}{0pt}
\begin{tabularx}{\textwidth}{@{}>{\raggedright\arraybackslash}p{3.9cm} X X@{}}
\toprule
\textbf{AIR element} & \textbf{ISO/IEC 27035 series} & \textbf{NIST SP 800-61 (rev. 3)} \\
\midrule
\textbf{Incident ID} & Incident number & \textbf{--} \\
\textbf{Incident Priority} & \textbf{--} & GV.RM-06 \\
\textbf{Incident Reporter} & Reporting person: Name $+$ Contact information & GV.RR-02 \\
\textbf{Incident Coordinator} & Point of contact (PoC): Name $+$ Contact information & \textbf{--} \\
\textbf{Detection Timestamp} & Date and time the incident was discovered & RS.MA-02 \\
\textbf{Detection Source} & What it occurred & \textbf{--} \\
\textbf{Timeline of Events} & \textbf{--} & RC.CO-03 \\
\textbf{Affected Assets/Systems} & Initial views on assets/components affected & DE.CM-02 \\
\textbf{Affected Dependencies} & Initial views on assets/components affected & DE.CM-02 \\
\textbf{Incident Type} & Incident category & RS.MA-03 \\
\textbf{Attack Vector} & How it occurred & RS.AN-03 \\
\textbf{Collected Evidence} & How it occurred & RS.AN-07 \\
\textbf{Incident Description} & How it occurred & RS.AN-03 \\
\textbf{Safety Impact} & Adverse business impact/effect of incident & \textbf{--} \\
\textbf{Operational Impact} & Adverse business impact/effect of incident & \textbf{--} \\
\textbf{Environmental Impact} & Adverse business impact/effect of incident & \textbf{--} \\
\textbf{Financial Impact} & Total recovery cost from incident & \textbf{--} \\
\textbf{Estimated Time to Safe Recovery} & \textbf{--} & Adapted from RC.RP-01 \\
\textbf{RTO Status} & Adapted from ISO7IEC 27002:2022 ICT readiness control (pp 48-49) & \textbf{--} \\
\textbf{Process Response Actions} & Actions taken to resolve incident & RS.MI-01 \\
\textbf{OT/IT Network Response Actions} & Actions taken to resolve incident & RS.MI-01 \\
\textbf{HSE Response Actions} & Actions taken to resolve & RS.MI-01 \\
\textbf{Regulatory Trigger} & \textbf{--} & GV.OC-03 \\
\textbf{Reporting Channels Used} & \textbf{--} & GV.RM-05 \\
\textbf{Internal and External Reporting} & Internal/External individuals/entities notified & RS.CO-02, RS.CO-03 \\
\bottomrule
\end{tabularx}
\endgroup
\begin{tablenotes}
\scriptsize
\item \textbf{Note:} ``\textbf{--}'' indicates no direct link in the cited standard.
\end{tablenotes}
\end{threeparttable}
\end{table}

A few AIR elements have no analogue in either IT standard. We adapted the RC.RP-01, from CSF 2.0, to create \textbf{Estimated Time to Safe Recovery}. This element adds an explicit forecast by the attacked organisation of when operations will stabilise. The coordinated efforts with external parties can adjust this forecast. This adaptation is crucial in OT environments where multiple impacts influence decision-making. Additionally, no parallel exists in ISO/IEC 27035 or NIST SP 800-61 (rev. 3) for \textbf{RTO Status}, so the element is derived from the ICT-readiness control in ISO/IEC 27002:2022 (pp 48–49), linking real-time progress to business-continuity targets. These additions serve to illustrate that, while IT standards provide a strong foundation, an OT‐focused framework must expand them to capture operational and safety realities, especially those in critical infrastructure.

\section{AIR exercises: embedding the framework in OT practice} \label{AIR_practice}

After outlining the characteristics of AIR, we designed three exercises to explore its implementation. The first Section (\ref{ex_ot_stand}) is a consistency exercise, where we revisit the OT standards and examine their original texts for statements that align with AIR elements. In this exercise, we highlight how the proposed elements facilitate the interpretation of the text of each standard. The second (\ref{ex_ot_adapt}) reimagines the OT frameworks discussed earlier. In this exercise, we revisit the original formulations, adding an ``AIR layer.'' The third Section (\ref{ex_case_study}) presents a case study, where we deploy AIR to analyse the 2015 Ukrainian Oblenergos Attack, exemplifying its operational potential.

\subsection{AIR compliance to OT standards} \label{ex_ot_stand}

The ISA/IEC 62443 series \cite[Part 2-1, Section 12.2.2]{IEC62443} recommends that asset owners establish policies and procedures for reporting security-related incidents in industrial automation and control systems. Because this guideline is generic, we link its supporting guidance to AIR via Part 3-3, FR-6 (timely response to events), which we map to specific AIR elements.

NIST SP 800-82 (rev. 3) \cite{NIST80082r3} likewise stresses prompt incident reporting and provides OT-specific context. It notes that the mechanisms enabling reporting do not need to be integrated into OT systems. Where integration exists, vendor neutrality is expected, and AIR adopts this design stance. Where mechanisms remain external, a standardised tool improves coordination among parties, which is aligned with AIR's primary objective. 

NERC CIP-008-6 \cite{NERCCIP} prescribes minimum reportable attributes for incidents (e.g., functional impact, attack vector, level of intrusion, and notification), sets notification deadlines for regulators, and specifies evidence-retention expectations. AIR encompasses all these factors.

Table \ref{tab:air_mapping_ot} shows how the original texts regarding the reporting procedure align with AIR elements. The mapping illustrates the operationalisation of AIR to facilitate the interpretation of the text of each standard, converting broad guidance into a structured and shareable format.

\begin{table}[ht]
\caption{Mapping of OT standards to AIR element(s)}
\label{tab:air_mapping_ot}
\footnotesize
\resizebox{\textwidth}{!}{
\begin{tabular}{p{2.5cm} p{10cm} p{4cm} p{4cm}}
\toprule
\textbf{Standard} & \textbf{Original Text} & \textbf{Text Excerpt} & \textbf{AIR Element(s)} \\
\midrule

\multirow{6}{*}{\makecell{ISA/IEC\\62443 series}} &
\multirow{6}{10cm}{
``Once detected, events should be reported to the \underline{\textit{appropriate personnel}} and assigned the \underline{\textit{appropriate priority}} for handling to ensure security risk targets are met. Reporting provides \underline{\textit{awareness of event occurrences}} and \underline{\textit{ensures that action can be taken as needed}} in a timely manner. Timeliness, therefore, is important and is a factor in \underline{\textit{reducing risks to a tolerable level}}. Events should be evaluated to determine \underline{\textit{who should receive notification}}, and if supported, their priority. Once that determination is made, the system should be configured to have the events reported appropriately.'' \cite[Part 2-1, sec 12.2.2.2]{IEC62443}} &
appropriate personnel & \textbf{Internal and External Reporting} \\
& & appropriate priority & \textbf{Incident Priority} \\
& & awareness of event occurrences & \textbf{Timeline of Events} \\
& & ensures that action can be taken as needed & \textbf{Incident Description} \\
& & reducing risks to a tolerable level & \textbf{Estimated Time to Safe Recovery} AND \textbf{RTO Status} \\
& & who should receive notification & \textbf{Internal and External Reporting} \\

\midrule
\multirow{3}{*}{\makecell{NIST SP\\800-82 (rev.~3)}} &
\multirow{3}{10cm}{
``OT Discussion: The organisation should report incidents on a timely basis. \underline{\textit{CISA collaborates with international and}} \underline{\textit{private-sector computer emergency response teams (CERTs)}} to share \underline{\textit{control systems-related security incidents}} and \underline{\textit{mitigation measures}}. Control Enhancement: (1) OT Discussion: The automated mechanisms used to support the incident reporting process are not necessarily part of or connected to the OT.'' \cite[p 257]{NIST80082r3}} &
CISA collaborates with international and private-sector computer emergency response teams (CERTs) & \textbf{Regulatory Trigger} AND \textbf{Internal and External Reporting} \\
& & control systems-related security incidents & \textbf{Incident Type} \\
& & mitigation measures & \textbf{Process Response Actions} AND \textbf{OT/IT Network Response Actions} AND \textbf{HSE Response Actions} \\

\midrule
\multirow{7}{*}{\makecell{NERC CIP\\series}} &
\multirow{7}{10cm}{
\begin{minipage}[t]{9.8cm}\footnotesize
\begin{itemize}\setlength\itemsep{1pt}
 \item \underline{\textit{Applicable Systems}}
 \item Requirements include ``4.1.1 The \underline{\textit{functional impact}}'', ``4.1.2 The \underline{\textit{attack vector used}}'', and ``4.1.3 The \underline{\textit{level of intrusion that was achieved or attempted}}''.
 \item Measures notes ``Examples of evidence may include, but are not limited to, \underline{\textit{dated documentation of initial notifications and updates}} \underline{\textit{to the E-ISAC and NCCIC}}''.
 \item Entities must notify regulators ``\underline{\textit{one hour after the determination of a Reportable Cyber}} \underline{\textit{Security Incident}}''. \cite[Table R4]{NERCCIP}
\end{itemize}
\end{minipage}} &
Applicable Systems & \textbf{Affected Assets/Systems} AND \textbf{Affected Dependencies} \\
& & functional impact & \textbf{Safety Impact} AND \textbf{Operational Impact} AND \textbf{Environmental Impact} \\
& & attack vector used & \textbf{Attack Vector} \\
& & level of intrusion that was achieved or attempted & \textbf{Incident Priority} AND \textbf{Incident Description} \\
& & dated documentation of initial notifications and updates & \textbf{Timeline of Events} \\
& & to the E-ISAC and NCCIC & \textbf{Internal and External Reporting} \\
& & one hour after the determination of a Reportable Cyber Security Incident & \textbf{Regulatory Trigger} \\

\bottomrule
\end{tabular}
}
\end{table}

\subsection{Addition of AIR to literature OT frameworks} \label{ex_ot_adapt}

\begin{figure}[ht]
  \centering
  \begin{subfigure}{0.45\textwidth}
    \centering
    \includegraphics[width=\linewidth]{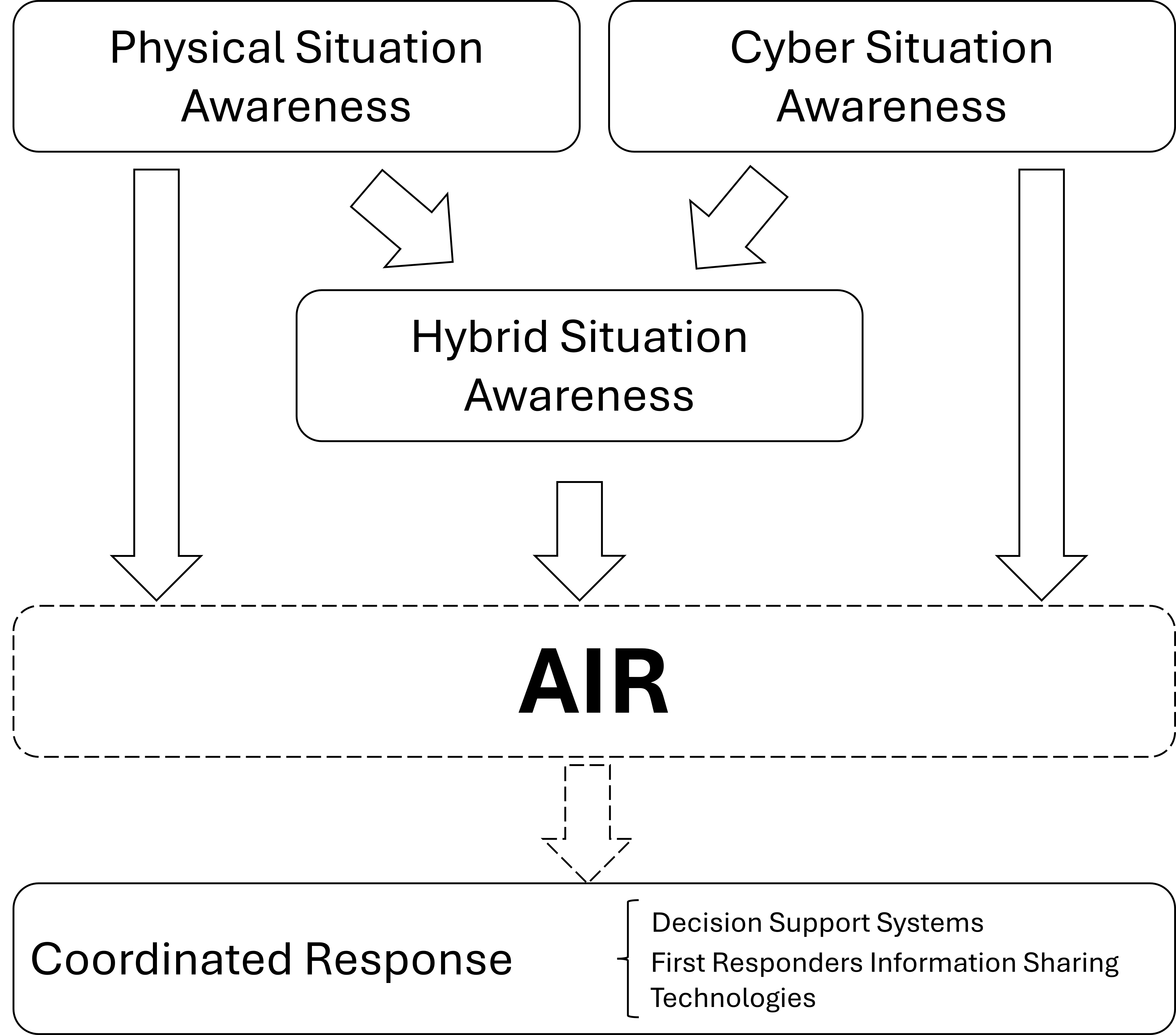}
    \caption{}
    \label{fig:air_praetorian}
  \end{subfigure}
  \hfill
  \begin{subfigure}{0.45\textwidth}
    \centering
    \includegraphics[width=\linewidth]{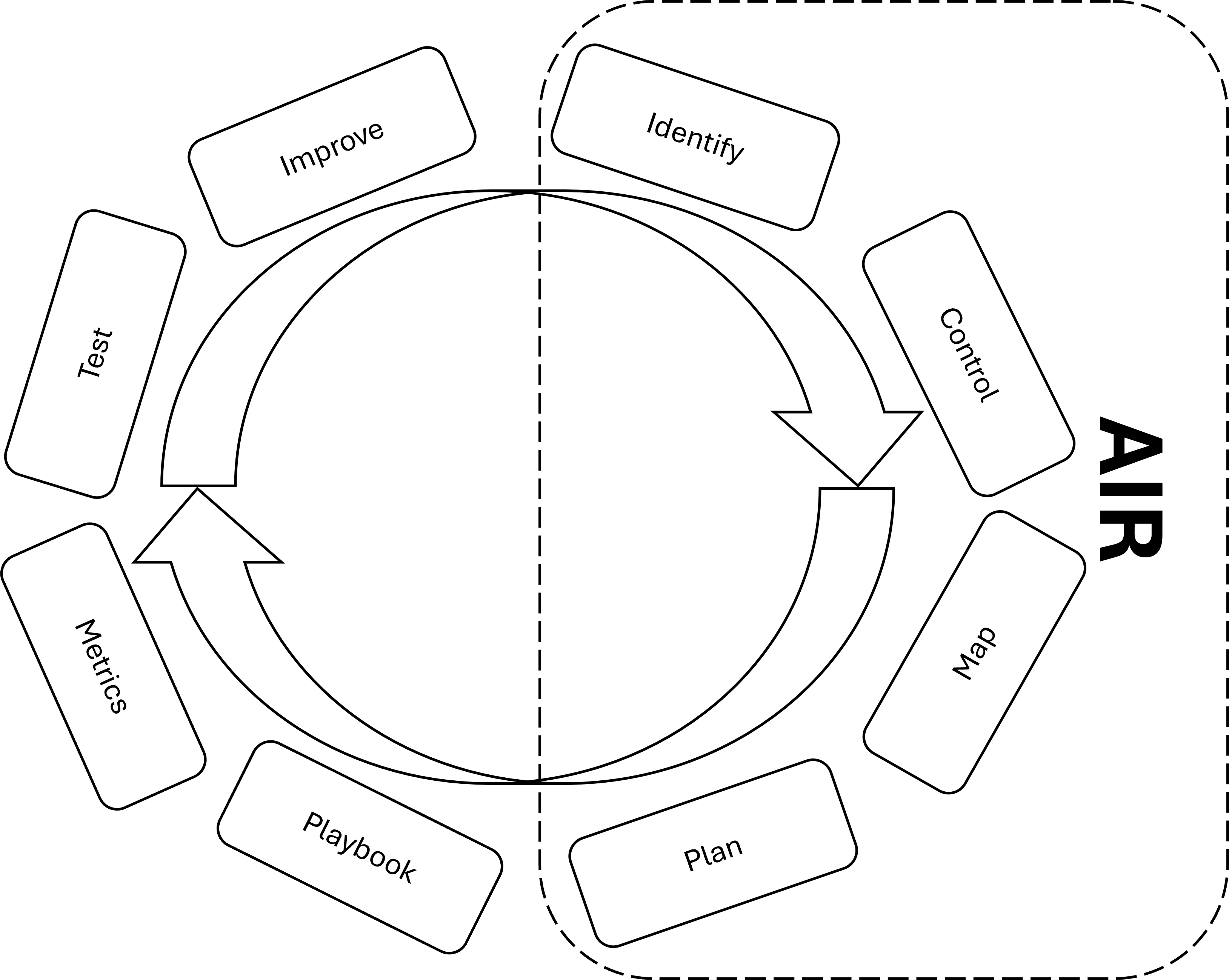}
    \caption{}
    \label{fig:air_crof}
  \end{subfigure}
    \caption{\texorpdfstring{AIR layer added to (a) PREATORIAN \cite{Papadopulos2024PRAETORIAN} and (b) CROF \cite{onwubiko2024recovery}.}{AIR layer added to (a) PREATORIAN and (b) CROF}}
  \label{fig:AIR_frameworks}
\end{figure}

AIR was not designed to replace any of the frameworks discussed in Section \ref{lit_fram}, but rather to be added as an additional layer, the “AIR layer,” depicted in Fig. \ref{fig:AIR_frameworks}. After this addition, AIR only activates when an OT framework escalates an event to an incident, typically via a severity metric. Delaying activation avoids effort for low-impact events and minimises computational load on the host framework. Once triggered, AIR runs in parallel with its host, feeding from pre-processed data, and is complete (shareable) by the time the OT framework reaches its termination point. This exercise represents the agnostic perspective of AIR, given its interconnection to existing frameworks.

POROS \cite{iaiani2021poros} flags case-specific attack mechanisms but needs supplementary documentation. At that point, practitioners can transfer POROS worksheet to AIR's standardised report, share it with peers and regulators, and request defensive guidance. In DSACR \cite{ma2024dsacr}, the Incident Detection and Analysis module records location, event type and risk level. When the score crosses the incident threshold, data is populated in AIR. The resulting report, with attack tactics, impact estimate, and recovery status, can be shared with stakeholders to make them aware and refine countermeasures. PRAETORIAN \cite{Papadopulos2024PRAETORIAN} comprises data from situational awareness modules, populating AIR. The final report would be shared with stakeholders (internal and external) in a standardised way, offering harmonisation and noise reduction across stakeholders. In CyberSentry \cite{Khadidos2025CyberSentry}, AIR is populated with the same data used by the original framework, and reports would be almost ready for cross-party disclosure after class prediction, since they would require estimated impacts. XA4AS \cite{seid2024xa4as} raises an alert when a predetermined optimisation goal fails. At that point, AIR packages the failure parameters and solicits external expertise, enabling the control loop to converge on a new global goal more quickly.

In TiDICS \cite{staves2024framework}, AIR is triggered when the framework is unable to map basic events to mitigation techniques. During the rules-of-engagement phase, the framework compiles data in accordance with ISO 31000. AIR then shares unresolved events with external specialists, speeds diagnosis, and notifies stakeholders. Some PRESeT \cite{schaeffer2014resilience} policies may halt operations (e.g. connection dropping) when challenge events exceed a threshold. AIR provides the channel for coarse-grained adaptation advice and fulfils any regulatory reporting triggered by the same event. CROF's \cite{onwubiko2024recovery} first four categories (Identify, Control, Map and Plan) supply nearly all AIR fields. Given this, AIR can be generated with minimal extra effort and used to coordinate recovery actions across internal and external stakeholders.

For each case, the ``AIR layer'' augments the host framework by standardising incident data and eases disclosure and coordination with stakeholders, without disrupting the framework's native workflow.

\subsection{AIR case study application: 2015 Ukrainian oblenergos attack} \label{ex_case_study}

\begin{sidewaysfigure}
  \centering
\includegraphics[width=\textheight]{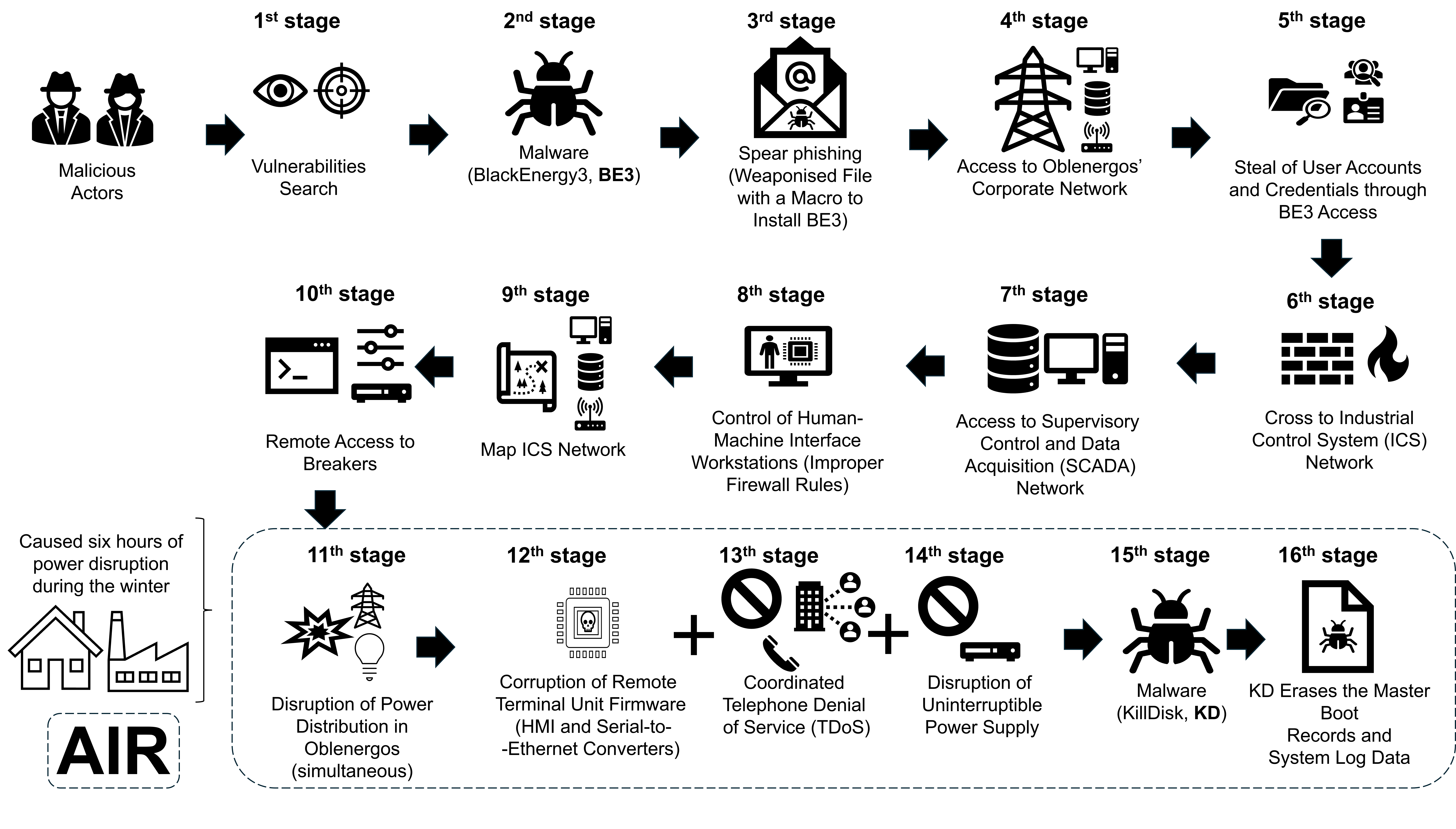}
  \caption{The 2015 Ukrainian oblenergos attack and the AIR framework}
  \label{fig:oblenergos}
\end{sidewaysfigure}

Before moving on to the case study, we present a visual representation (Fig. \ref{fig:oblenergos}) of all the steps involved in the 2015 Ukrainian oblenergos attack. The dashed part will be discussed further in the following paragraphs, given that it represents the stages where response teams could have employed AIR.

On 23 December 2015, a cyberattack targeted three Ukrainian regional distribution companies (oblenergos): Prykarpattyaoblenergo, Kyivoblenergo and Chernivtsioblenergo. Approximately 225,000 customers, including households and businesses, lost power for up to six hours in winter \citep{Booz2019,SANS2016,Whitehead2017}.

Following months of preparation, the attackers harvested credentials via spear-phishing using BlackEnergy3 and accessed the SCADA network through VPN and remote tools \citep{NCCIC-ICS-alert,Shehod2016}. At about 3:30 p.m. local time, operators observed unauthorised actions on SCADA human–machine interfaces (HMIs) as commands were sent to open approximately 30 breakers at dozens of substations, interrupting power distribution \citep{NCCIC-ICS-article,Booz2019,Whitehead2017}. Malicious firmware was pushed to serial-to-Ethernet converters that bridge SCADA and field equipment, which blocked remote re-closure and forced manual restoration \citep{Whitehead2017,Shehod2016}.

The response was complicated by a telephony denial-of-service (TDoS) campaign that flooded utility call centres with automated calls, blocking lines and reducing situational awareness \citep{Booz2019,Shehod2016,NCCIC-ICS-article}. In parallel, unauthorised UPS shutdowns were issued via remote management interfaces, affecting systems that supported control and communications, and internal monitoring platforms lost visibility \citep{NCCIC-ICS-alert}. In at least one instance, a system reboot activated KillDisk, which corrupted the master boot record (MBR) and erased data on operator workstations and SCADA servers, delaying restoration \citep{NCCIC-ICS-article,Shehod2016}. Field crews carried out on-site switching at affected substations, which remained offline until manual restoration that took up to six hours \citep{NCCIC-ICS-article,Whitehead2017}.

We apply AIR to collect official information, assuming it was available during the incident. This exercise is shown in Table \ref{tab:air_case_study}. Information is incomplete (--) because we rely solely on official reports and peer-reviewed papers. Even so, AIR captures vital information with the potential to support collaboration and to provide timely insights to operational and support teams. It gives managers and regulators sufficient visibility. Shareable information from the report would be available to affected parties.

\begin{table}[!h]
\centering
\scriptsize
\caption{AIR framework applied to the 2015 Ukrainian oblenergos cyberattack}
\label{tab:air_case_study}
\begin{threeparttable}
\setlength{\tabcolsep}{4pt}
\renewcommand{\arraystretch}{1.07}
\begin{tabularx}{\textwidth}{@{}>{\raggedright\arraybackslash}p{3.4cm} >{\raggedright\arraybackslash}X@{}}
\toprule
\textbf{AIR element} & \textbf{Information} \\
\midrule
\textbf{Incident ID} & \textbf{--} \\
\textbf{Incident Priority} & Orange (High, 10), following Table B.6 \cite{IEC62443} \\
\textbf{Incident Reporter} & \textbf{--} \\
\textbf{Incident Coordinator} & \textbf{--} \\
\textbf{Detection Timestamp} & 2015-12-23 13:30 UTC (15:30 local time) \\
\textbf{Detection Source} & Utility operator observations of unauthorised HMI actions \\
\textbf{Timeline of Events} &
1) 13:30 UTC (15:30 local): breakers opened via SCADA HMI (\(\sim\)30);
2) Firmware pushed to serial-to-Ethernet converters, blocking remote re-closure;
3) TDoS floods utility call centres;
4) UPS shutdowns issued via remote management interfaces;
5) KillDisk activates after reboot (at least one system);
6) Manual restoration by field crews; service restored within up to six hours \\
\textbf{Affected Assets/Systems} & SCADA HMIs, circuit breakers, serial-to-Ethernet converters, UPS units, operator workstations \\
\textbf{Affected Dependencies} & Call-centre phone lines; communications hardware supporting control systems; internal monitoring platforms \\
\textbf{Incident Type} & Threat event: manipulation of control (Table 21 \cite{NIST80082r3}); incident: malicious and direct \cite[p.~180]{NIST80082r3} \\
\textbf{Attack Vector} & Spear phishing (to deploy BlackEnergy3) leading to credential theft and SCADA network access \\
\textbf{Incident Description} & Attackers operated SCADA HMIs to open \(\sim\)30 breakers; malicious firmware on serial-to-Ethernet converters blocked remote re-closure; TDoS and UPS shutdowns impeded coordination; manual restoration required. \\
\textbf{Collected Evidence} & HMI and remote-access logs; evidence of malicious firmware on serial-to-Ethernet converters; UPS management logs showing shutdowns; forensic images indicating KillDisk on operator workstations/SCADA servers; malware artefacts (BlackEnergy3, KillDisk). \\
\textbf{Safety Impact} & \textbf{--} \\
\textbf{Operational Impact} & Power outage for \(\sim\)225{,}000 customers; SCADA and substations offline \\
\textbf{Environmental Impact} & \textbf{--} \\
\textbf{Financial Impact} & \textbf{--} \\
\textbf{Estimated Time to Safe Recovery} & \textbf{--} \\
\textbf{RTO Status} & \textbf{--} \\
\textbf{Process Response Actions} & Manual breaker closures performed by field crews \\
\textbf{OT/IT Network Response Actions} & \textbf{--} \\
\textbf{HSE Response Actions} & \textbf{--} \\
\textbf{Regulatory Trigger} & \textbf{--} \\
\textbf{Reporting Channels Used} & Internal plant reporting; escalation to Ukrainian CERT \\
\textbf{Internal and External Reporting} & Ukrainian CERT \\
\bottomrule
\end{tabularx}
\begin{tablenotes}
\scriptsize
\item \textbf{Note:} \textbf{--} indicates information not available or unknown. “TDoS” denotes telephony denial of service.
\end{tablenotes}
\end{threeparttable}
\end{table}

\section{Concluding remarks, future studies and limitations} \label{final}

This paper aimed at addressing the absence of a structured and standardised framework to ongoing incident reporting, a critical gap in the OT cybersecurity landscape. While OT standards establish requirements for incident reporting and forensic evidence preservation, they do not specify data elements to collect during an incident. The 2015 Ukrainian power grid attack demonstrated that the lack of unified information flows can paralyse response efforts, prolong recovery and impact customers and businesses. In contrast, IT standards provide guidelines for incident reporting but fall short in accounting for the technical and organisational specificities of OT environments.

We proposed AIR, an Agnostic Incident Reporting framework to address this gap. AIR is a reporting tool designed specifically for OT cybersecurity incidents. Our proposed framework can be added to existing OT frameworks and convert broad guidance into a structured and shareable format, embedding compliance. AIR adapts established IT reporting principles to meet the operational and regulatory demands of OT systems. It introduces a modular structure comprising 25 elements organised into seven thematic fields. These elements document who is responsible, when events occur, where the compromise takes place, how the intrusion unfolds, what mitigation actions are taken, and the expected or observed impacts. Additionally, AIR facilitates coordination among technical, managerial, and regulatory stakeholders.

We designed exercises for AIR targeting three dimensions: compliance with OT standards, integration with OT frameworks, and applicability to incidents. The first exercise demonstrated how AIR operationalises abstract language from high-level standards into concrete reporting elements. The second exercise displayed that AIR can operate across multiple OT frameworks, with defined activation points and interfaces. The third exercise applied AIR to the 2015 Ukrainian oblenergos attack as a case study to test whether structured reporting would support situational awareness, faster containment, and transparent communication with affected parties.

Regardless the organisational cybersecurity maturity, AIR harmonises data exchange, reduces miscommunication during crises, and meets regulatory and contractual obligations. AIR provides a scalable foundation for future efforts to standardise ongoing incident reporting across the OT domain.

Our study is not without limitations. First, although the information elements that constitute AIR were informed by existing literature and standards, they were not validated through consultation with OT practitioners. As with many proposed frameworks, AIR remains a conceptual contribution until further validated in practice. Some information fields may also be considered too sensitive for sharing, even under SLAs or NDAs. Second, the case study applied AIR retrospectively to a completed incident rather than in an ongoing operational setting. This decision reflects the ethical and safety risks of testing a new framework during an incident. Finally, specific information fields in our case study table remain unpopulated due to the limited availability of verified public data. This limitation is not unique to the Ukrainian incident, since selecting a different case would not eliminate this constraint, as reliable real-time data from OT incidents is generally scarce.

These limitations inform our suggestions for future research. Our next step is to validate AIR's reporting elements with domain experts. Following validation by experts, AIR can be implemented as an add-on or plug-in within existing OT security tools and tested under experimental conditions. Insights from such trials will inform refinements to the framework. Ultimately, AIR can be deployed in ongoing OT cybersecurity incidents, provided the experimental phase confirms its operational readiness and value. 

\section*{Acknowledgments} \label{ack}
This study was financed by UK Research and Innovation and the Welsh Government (KTP project reference 14104). The authors thank Dr Michael Robinson, Cybersecurity Research Engineer at Airbus, and Angela Smith, Head of Cyber Innovation at Airbus, for the constructive criticism of the previous versions of the manuscript.

\section*{Declaration of Competing Interest } \label{comp}
The authors of this manuscript have no competing or non-financial interests, currently or previously, including serving in an editorial capacity for the journal we are submitting to.

\section*{Data Availability} \label{data}
Data sharing does not apply to this article as no new data were created or analysed in this study.

\section*{CRediT Taxonomy} \label{credit}
\textbf{Nubio Vidal} -- Conceptualisation, Formal analysis, Investigation, Methodology, Visualisation, Writing – original draft, Writing – review \& editing. \textbf{Naghmeh Moradpoor} -- Conceptualisation, Formal analysis, Funding acquisition, Project administration, Supervision, Validation, Writing – original draft, Writing – review \& editing, Resources. \textbf{Leandros Maglaras} -- Validation, Writing – review \& editing

%\subsubsection{\discintname}
%The authors have no competing interests to declare relevant to this article's content.

%\end{credits}

\bibliographystyle{elsarticle-num.bst}
\bibliography{AIR_ref}

\end{document}